\documentclass{scrartcl}

\usepackage{csquotes}

\usepackage{amsmath}
\allowdisplaybreaks
\usepackage{amsfonts}
\usepackage{amsthm}

\usepackage{booktabs}
\usepackage{paralist}
\usepackage{comment}

\usepackage{url}

\usepackage{notes}

\usepackage{graphicx}

\usepackage{xcolor}

\usepackage{listings}
\lstdefinelanguage{sparql}{
	morecomment=[l][\color{olivegreen}]{\#},
	morestring=[b][\color{blue}]\",	morekeywords={SELECT,CONSTRUCT,DESCRIBE,ASK,WHERE,FROM,NAMED,PREFIX,BASE,OPTIONAL,FILTER,GRAPH,LIMIT,OFFSET,SERVICE,UNION,EXISTS,NOT,BINDINGS,MINUS,a},
	sensitive=true
	}
\usepackage{algorithm}
\usepackage{algorithmicx}
\usepackage[noend]{algpseudocode}
\algdef{SE}[DOWHILE]{Do}{DoWhile}{\algorithmicdo}[1]{\algorithmicwhile\ #1}%

\usepackage{mathtools}

\usepackage{listings}
\usepackage{chngcntr}

\usepackage{balance}

\usepackage{authblk}

\usepackage{pdflscape}

\newtheorem{definition}{Definition}

\newtheorem{example}{Example}

\providecommand{\keywords}[1]{\textbf{\textbf{Keywords:}} #1}

\begin{document}

\title{FedMark: A Marketplace for Federated Data on the Web}

\author[1,2]{Tobias~Grubenmann}
\author[1]{Abraham~Bernstein}
\author[1]{Dmitry~Moor}
\author[1]{Sven~Seuken}
\affil[1]{Department of Informatics, University of Zurich}
\affil[2]{Department of Computer Science, The University of Hong Kong}

\date{}

\maketitle

\begin{abstract}
\noindent
The Web of Data (WoD) has experienced a phenomenal growth in the past. This growth is mainly fueled by tireless volunteers, government subsidies, and open data legislations.
The majority of commercial data has not made the transition to the WoD, yet. The problem is that it is not clear how publishers of commercial data can monetize their data in this new setting. Advertisement, which is one of the main financial engines of the World Wide Web, cannot be applied to the Web of Data as such unwanted data can easily be filtered out, automatically.
This raises the question how the WoD can (i) maintain its grow when subsidies disappear and (ii) give commercial data providers financial incentives to share their wealth of data.  
In this paper, we propose a marketplace for the WoD as a solution for this data monetization problem. Our approach allows a customer to transparently buy data from a combination of different providers. To that end, we introduce two different approaches for deciding which data elements to buy and compare their performance. We also introduce FedMark, a prototypical implementation of our marketplace that represents a first step towards an economically viable WoD beyond subsidies.
\end{abstract}


\section{Introduction}\label{sec:introduction}

Inspired by the WWW's characteristics, the \emph{Web of Data (WoD)} is a decentralized repository of data, where many data tenants publish and manage interlinked datasets whether indexed or not. Its goal is to provide a globally distributed knowledge base where query engines can mix-and-match data from various, distributed data-sources towards answering queries. Most current datasets on the WoD are freely available, either subsidized by governments via data access laws and research grants or maintained by enthusiasts. Some provider of datasets will be able to maintain funding for their datasets and continue to be open. Right now, however, only a third of the public SPARQL endpoints have an uptime of 99\% and above~\cite{builAranda2013}. Without financial incentives, many promising datasets will be poorly maintained or unavailable as there is no one willing to invest time and money to keep the data up-to-date and the endpoint running~\cite{ErikBrynjolfsson}. 

Unfortunately, most incentive mechanisms from the WWW do not translate to the WoD, as data is queried by machines rather than humans. Consequently, WoD query results often do not contain any attribution to the original source and algorithms can simply filter out any contained advertisement removing most non-monetary benefits from the publisher. Hence, the many motivations typically entailed in authoring a web page---fame, money through advertisement, acknowledgment, or recognition---do not carry over to the WoD. Even though provenance techniques exist, such meta-information will not be shown to the user if not explicitly requested.

\begin{example}\label{ex:ipa_introduction}
	Consider an Intelligent Personal Assistant (IPA)---a computer program assisting a user by automatically searching the WoD for relevant information and interacting with other computer programs---which searches the WoD on a daily basis for relevant information regarding error messages that occur while working with computers (akin to paring osquery\footnote{\url{https://osquery.io}} with an error recognition and suggestions database extracted from stackoverflow). The IPA uses context information about the program throwing the error, the operating system, and other relevant data to find articles, comments, and other pieces of information that can help understanding and solving the problem which caused the error. Unlike a keyword-based search in the WWW, the IPA could find information that is much more specific to the context of the error and provide different suggestions without any interaction needed by the user.
	
	IPAs like the one presented above are one of the big promises of the Web of Data \cite{bernerslee2001semantic}. However, the question we raise in this paper is how people can be motivated to create the content needed to enable such a vision. In our example, many of the articles, comments, etc. about problems are written by fellow users, who already encountered the problem and are now sharing the gathered knowledge. In the WWW, such fellow users are credited when giving a helpful answer and are usually thanked by others. The website hosting the platform for this knowledge exchange makes money by showing advertisement and job offers to the users. As we can see, there are two \emph{incentive mechanisms} which keep the knowledge exchange platform alive:
	\begin{inparaenum}[(1)]
		\item \emph{acknowledgement} and \emph{reputation} incentivize users to share their knowledge and
		\item \emph{money} from advertisements and job offers help to finance the platform.
	\end{inparaenum}
	
	In our IPA scenario, however, the users creating the knowledge and the platform offering the knowledge are transparent to the end-user who is consuming the knowledge through the IPA. In fact, the lack of any end-user interaction required is one of the big advantages of having an IPA in the first place. But how can such a scenario work if we remove the incentive mechanisms which were used in the WWW to create the necessary knowledge?
\end{example}

Due to the above-mentioned reasons, we believe that in the long-term many providers of semantic data have to charge fees (directly or indirectly) for accessing their data in order to finance their services. However, as soon as users are charged for the data they consume, economic considerations play an important role during query planning. In particular, the users (or programs acting on behalf of the users) have to decide to which data sources they buy the right for accessing. This decision, in turn, depends on how much the bought data can contribute to a specific query. As we showed in \cite{grubenmann2017-ISWC}, it is very difficult to decide \emph{before} query execution how much a certain source can contribute to a query answer. However, \emph{after} query execution it is too late to decide against the inclusion of some sources, as the data is already bought. Whilst data synopses can sometimes help in deciding which data sources might be worth accessing for a specific query, our analysis showed that there is no universal approximation method which can consistently yield good enough results to judge the economic utility of a source for a specific query. \emph{These findings question the practicability of a scenario, where data providers charge customers directly for accessing their data.}

Alternatively, one might argue, the nature of data will lead to natural monopolies and we should concentrate on building \emph{one large centralized database}. Such a database would allow the maintainer to extract monopolistic fees for its usage, which could pay for the data maintenance. For example, data services such as \emph{Bloomberg}, \emph{LexisNexis}, or \emph{Thomson Reuters} charge customers high fees for accessing their data primarily using a subscription-based model. These sellers can price their services by calculating a quasi-monopolistic price on their whole data offering \cite{Bakos1999}.
Indeed, most non-monopolistic settings struggle to find a good pricing-scheme. The \emph{Azure DataMarketplace} \cite{azure}, e.g., closed in March 2017, due to the lack of attraction. The Copenhagen City Data Exchange is still trying to figure out how to find a good way to price their data sets.\footnote{\url{https://www.citydataexchange.com/}, personal communication} 
However, none of these solutions provide their data in a way such that they can be queried in a federated fashion. They do not provide the means to join datasets from multiple providers and access can only be purchased in an all or nothing approach, thereby forgoing the complementarities the WoD would enable. This is a serious drawback, because customers are often interested in a specific \emph{combination} of data from different providers that are joined in a certain way.
Also, as Van Alstyne et al. \cite{ErikBrynjolfsson} argue, the incentive misalignments in a federated system based on these principles may lead to significant data quality problems.
Finally, some users may not be prepared to pay for the large bundles of data sold by these monopolists as they are only interested in occasional or very partial access. These are left out of these markets.
Hence, the central question of this paper is \emph{how can we facilitate a financial sustainable and decentralized WoD without government subsidies or federation-averse centralization and fulfill the promise of the data economy~\cite{Economist2017}?} 

This paper proposes \textsf{FedMark}, \textbf{a marketplace for data following the WoD principles of federated querying}. In contrast to the settings described above, \textsf{FedMark} allows a user (or customer) to submit a query and decide \emph{after} query execution which data should be bought without accessing the data (and incurring the moral hazard of not wanting to pay for already seen data).
To this goal, \textsf{FedMark} acts as a mediator between the customer and various data providers. \textsf{FedMark} executes the query but does not pass the query answer to the customer, yet. Based on a \emph{summary} of the full query answer, the customer can decide which parts of the query answer to buy. This selection of a subset of the query answer---which we call an \emph{allocation}---can be done either manually by composing the query answer based on personal preference or automatically by using an \emph{allocation rule}, which automatically determines how the query answer should be composed based on the available information. As manual composition is impractical in most large settings, allocation rules are crucial tools to deal with the exponentially growing number of possible allocations.

\textsf{FedMark} introduces a new paradigm towards querying and pricing the WoD relying on a market-based approach and principles of federated querying. This paradigm enables data providers to finance their wealth of data without relying on subsidies but on per-query fees. Our approach has the following advantages:
\begin{itemize}
	\item A customer can buy a query answer from a \emph{combination} of different data providers in a transparent way.
	\item Given a customer's query, our marketplace creates a query answer based on all available datasets from which a customer can allocate his most preferred subset.
	\item The price for the allocation depends only on the data \emph{contributing} to the allocation. Especially, the price is independent of query execution (in particular, the order of joins) and size of underlying datasets. Hence, \textsf{FedMark} compensates data providers for the value of the data they contribute for the specific query answer.
\end{itemize}

\noindent Our contributions are, hence:
\begin{itemize}
  \item the introduction of a market-based paradigm towards querying and pricing,
  \item the presentation of two different allocation rules for such a marketplace,
  \item the introduction of a prototype system \textsf{FedMark} implementing this paradigm, and
  \item the thorough evaluation establishing the practicality of our approach in terms of run-time overhead and utility maximization.
\end{itemize}

In the following, we start with some preliminaries about the Web of Data. We continue with the discussion of related work and then introduce our data market concept. This leads the way to our prototype implementation \textsf{FedMark} and the introduction of two allocation rules. Next, we perform an empirical evaluation of the runtime of the introduced allocation rules. We close with a discussion of the limitations of this study and an outlook for future work.

\section{Preliminaries}

\begin{figure}
	\centering
	\includegraphics[trim=0 20 0 20, clip, width=0.8\linewidth]{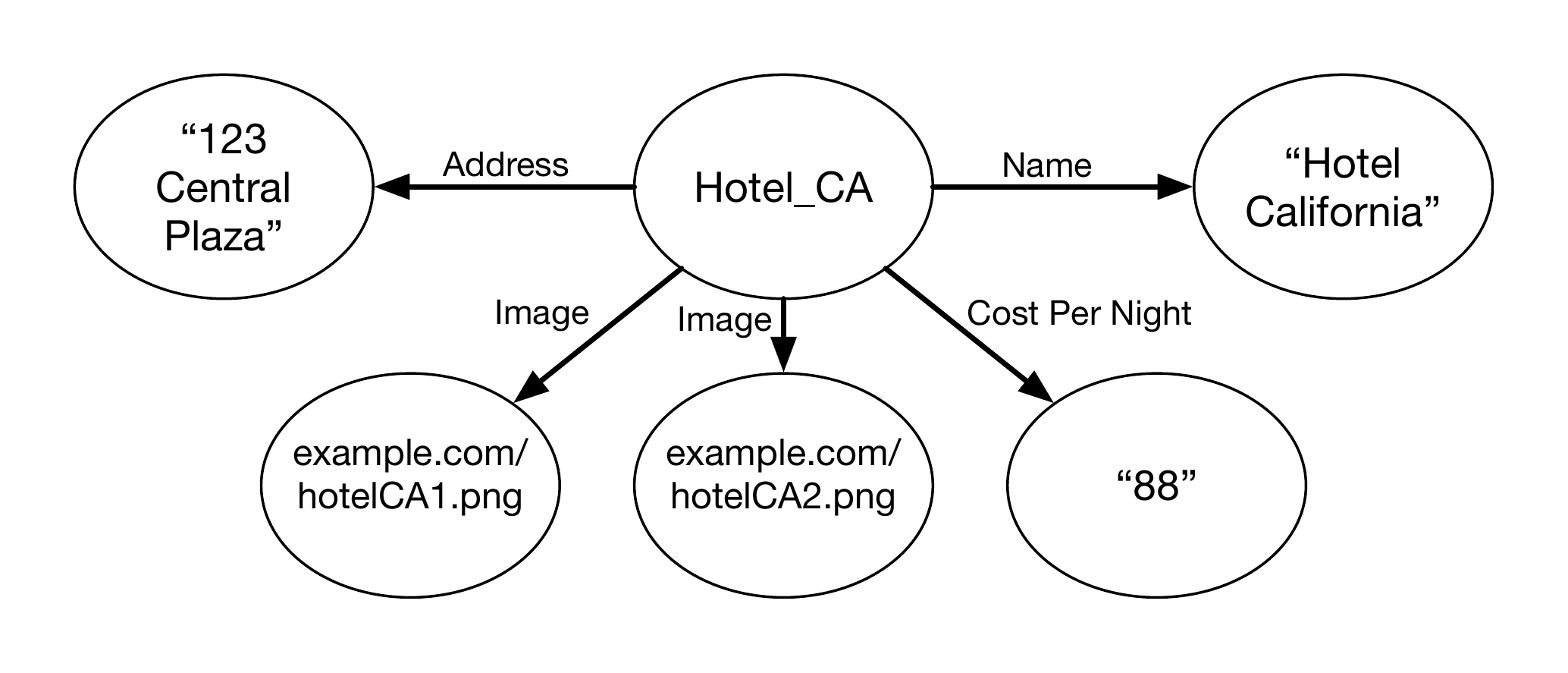}
	\caption{An example RDF graph describing a hotel.}\label{fig:RDFGraph}
\end{figure}

In the Web of Data, relations between resources are modelled using the Resource Description Framework (RDF) \cite{rdfW3C}, using \emph{resources} and \emph{literals}. A resource can denote anything, e.g., a website, an image, a physical thing, or an abstract entity. A literal is a string of characters with an optional datatype associated to it. The relations between resources and literals are modelled as statements consisting of a subject, object, and predicate linking the former two. A statement in RDF is also called a \emph{triple}. Fig. \ref{fig:RDFGraph} shows an example of how different information about a hotel can be modelled using such triples. Subjects and objects are illustrated with circles, predicates with arrows pointing from subjects to objects. Labels with quotation marks indicate literals, other labels indicate resource identifiers.

Users interested in the data represented in an RDF graph can use the query language SPARQL \cite{SparqlW3C}. Listing \ref{lst:SPARQLQuery} shows an example of how a user could ask for images of a hotel named "Hotel California". Identifiers starting with a "?" indicate variables. Each line ending with a "." indicates a \emph{triple pattern}, which can be matched to triples inside a graph. Such triple patterns can be used to form more complex \emph{graph patterns}, which can be combined with other operators like filter expressions, joins, and unions. The answer to a SPARQL query, which we refer to as \emph{query answer}, consists of bindings to the variables of the graph pattern after the \textsf{WHERE} clause projected to the variables specified after the \textsf{SELECT} clause. A \emph{SPARQL endpoint} is a web address which accepts SPARQL queries and returns query answers based on the RDF graph stored in the back-end.

Table \ref{tab:queryResult} shows the bindings which would be returned as query answer if the SPARQL query from Listing \ref{lst:SPARQLQuery} would be executed against the RDF graph from Fig. \ref{fig:RDFGraph}. Each row in the table represents a single \emph{solution mapping} to the query. A solution mapping is one possible set of bindings of variables to resources or literals which correspond to the queried RDF graph. A query answer is a set of such solution mappings. A query answer can contain one, multiple, or no solution mappings, depending on the RDF graph against which the query was executed \cite{SparqlW3C}.

\begin{lstlisting}[float=htb, caption={A query which asks for images for a hotel named \enquote{Hotel California}.}, label={lst:SPARQLQuery}, deletekeywords={ORDER, ORDERED}, captionpos=t]
PREFIX ex: <http://example.com/>
SELECT ?image
WHERE {
	?hotel ex:depicts ?image .
	?hotel ex:name "Hotel California" . }
\end{lstlisting}

\begin{table}[htb]
\centering
\caption{Result of the query in Listing \ref{lst:SPARQLQuery}.}\label{tab:queryResult}
	\begin{tabular}{llll}
		\hline
		?image \\ \hline
		example.com/hotelCA1.png \\
		example.com/hotelCA2.png \\ \hline
	\end{tabular}
\end{table}

It is possible to execute a query against multiple RDF graphs. Different RDF graphs can be made available on a single SPARQL endpoint or on different endpoints. In the latter case, a SPARQL query must be split up into subqueries which must be executed on the different servers. In this case, the different endpoint can be combined into a \emph{federation} of SPARQL endpoints. The (partial) query answers returned from the different machines must be joined together to form the final query answer.

\section{Related Work}

Our approach is based on standardized federated querying on the WoD \cite{buil-aranda2013}. As such, it relies on basic techniques of SPARQL querying \cite{perez2009}. Here, we very succinctly discuss the most recent federated querying techniques before elaborating on previous attempts of pairing market-based ideas in data management. 

\noindent
\textbf{Federated Querying on the WoD:} The traditional concepts for federated RDF querying provided integrated access to distributed RDF sources controlled by the query engine \cite{Harth2007,Quilitz,Virtuoso2006}. The drawback of these solutions is that they assume total control over the data distributions---an unrealistic assumption in the Web. Addressing this drawback, systems were proposed that do not assume fine-grained control: some exploit perfect knowledge about the \texttt{rdf:type} predicate distribution \cite{Langegger2008} while others proposed to extend SPARQL with explicit instructions controlling where to execute sub-queries \cite{Zemanek2007}. Often, however, the query writer has no ex-ante knowledge of the data distribution.

SPLENDID \cite{Gorlitz2011} proposed to exploit service descriptions and VoID statistics about each endpoint, to perform source selection and query optimization. HiBISCuS \cite{saleem2014}, on the other hand, maintains an index of authorities for certain URIs. FedX \cite{SchwarteHHSS11} uses no knowledge about mappings or statistics about concepts/predicates. It consults all endpoints to determine if a predicate can be answered (caching this information for the future). Fed-DSATUR \cite{vidal2016} is an algorithm for SPARQL query decomposition in federated settings without relying on statistics, indices, or estimates for source selection.
Forgoing any ex-ante knowledge about data sources and any requirements on data storage, Avalanche \cite{Basca2014} proposes an approach that combines data-source exploration followed by extensive parallelized and interleaved planning and execution.

Following another avenue, Hartig et al. \cite{Hartig2009a} describe an approach for executing SPARQL queries over Linked Open Data (LoD) based on graph search. LoD rules, however, require them to place the data on the URI-referenced servers---a limiting assumption, e.g., when caching/copying data.

Whilst these approaches provide a solid foundation for federated WoD querying, none of them considers the economic viability of their proposed solutions. Hence, we will extend this foundation with a market-based allocation approach to ensure economic viability. 

\noindent 
\textbf{Market-based Approaches towards Resource Allocation in Computational Systems:} 
The idea to use markets to allocate computational resources is almost as old as computers. Already in the 1960s, researchers used an auction-like method to determine who gets access to a PDP-1, the world's first interactive, commercial computer \cite{Sutherland1968FuturesMarketInComputerTime}. Since then, many market-based approaches for computational systems have been proposed.

Early research on market-based scheduling focused on the efficiency of computational resource allocation. The Enterprise system \cite{Malone1983} introduced a market for computational tasks. It efficiently allocated the tasks to multiple LAN-connected nodes, where task processors broadcast requests for bids and bid on tasks. Likewise, Spawn \cite{Waldspurger1992} utilized a market mechanism to optimize the use of idle resources in a network of workstations. More recently, \cite{Lai2005} proposed Tycoon, a distributed computation cluster, featuring a resource allocation model. The authors claim that an economic mechanism is vital for large scale resource allocation---a common problem on the Web. Furthermore, \cite{AlvinAuyoung} demonstrates how profit-aware algorithms outperform non-profit aware schedulers across a broad range of scenarios.

In data processing centric scenarios, \cite{Labrinidis2006} applied market-based optimizations to real-time query answering systems. \cite{Mariposa} proposed a WAN-scale Relational Database Management System with a market-based optimizer instead of a traditional cost-based one. \cite{dash} proposed a market-based approach for cloud cache optimization taking into account a user's value for getting an answer to a query. However, their approach focuses on the cost-side of cloud computing.

For relational databases, markets for SQL queries were proposed which sell data instead of computational resources for answering queries and use arbitrage-free pricing schemes to calculate payments \cite{deep2017,koutris:2013}. \cite{wang2016} proposed an auction mechanism for data which considers the negative externalities of allocating data to different buyers. However, they do not consider the possibility of joining datasets from different providers, which is an important aspect of the scenario we are investigating. To the best of our knowledge, none of these systems considers partial answers due to budget constraints or the possibly differing valuations of various users/queries, which is very typical in the WoD.

As a precursor to our research, we conducted a pilot study simulating a market platform for the WoD \cite{ZollingerEtAl}. This paper here represents a significant rework of the old pilot as it proposes a complete model, an improved market analysis, and a prototype implementation instead of a simulation. In \cite{moor2015}, we introduced the idea of using a double-auction for the WoD and showed the deficiency of the threshold rule in this setting together with three ways to correct them. However, our approach assumed that we have access to accurate join-estimates to produce satisfying results -- an assumption which might be hard to enforce in the WoD. In \cite{GrubenmannEtAl:DeSemWeb2017}, we presented our vision of a marketplace which allows customers to buy data from decentralized sellers in an integrated way. In this paper, we fulfil this vision and present our implementation of a federated marketplace for such decentralized data.

\section{Market Concept}\label{sec:concept}

We begin to describe our market concept by continuing Example \ref{ex:ipa_introduction} from Section \ref{sec:introduction}. Throughout this paper, we will extend this example to show how to decide which solution mappings to include in the customer's query answer and how much the customer must pay for it.

\begin{example}\label{ex:ipa_query}
	Consider a user who encounters the error message \texttt{"0x12345678"} while working with the (fictive) Integrated Development Environment (IDE) \enquote{Hack IDE} on a Java program. The IPA will recognize that the problem occurred and search for suggestions to fix the problem related to the error message. The IPA will order the different suggestions to the problem by their success rate. The IPA will use other information available, like the operating system, to refine the search and to get only relevant suggestions. Listing \ref{lst:errorQuery} shows how a SPARQL query generated by the IPA might look light. Each row of the query answer represents one possible suggestion to the problem with the corresponding success rate.
	
	We assume that at least part of the data needed to answer the query requires a payment from the user. In order to autonomously retrieve the query answer, the IPA buys the required data on behalf of the user in our marketplace. The marketplace finds the providers that offer datasets to answer this query. There might be multiple combinations of providers that would yield a non-empty query answer. Some of them might provide only suggestions without success ratings; others might provide only success ratings for suggestions, and some might provide both. As a result, there are multiple different combinations of datasets which produce (possibly) different query answers. Some of the query answers may contain only a few suggestions and ratings, whereas others may contain many, or none.
\end{example}

\begin{lstlisting}[caption={A SPARQL query asking an IPA can use to retrieve suggestions to a problem indicated by an error message.}, label={lst:errorQuery}, captionpos=t]
PREFIX ex: <http://example.com/>
SELECT ?suggestion ?rate WHERE {
  ?suggestion ex:success_rate ?rate .
  ?suggestion ex:err_code "0x12345678" .
  ?suggestion ex:program ex:hack_ide .
  ?suggestion ex:language ex:java .
  ?suggestion ex:os ex:os_x .
} ORDER BY DESC(?rate)
\end{lstlisting}

At the core of \textsf{FedMark} lies the ability for a customer to join data from different providers to buy solution mappings to a given query. Instead of buying all the solution mappings contained in a query answer, \textsf{FedMark} allows a customer to select a subset of the solution mappings---which we call an \emph{allocation}---and only paying the price of the allocated solution mappings.

Note that an allocation is also a query answer. We will refer to the result of the query execution as \emph{query answer} and the result of the allocation process as \emph{allocation} to emphasize the difference.

For a specific query, different combinations of providers' data might result in different (even empty) query answers. Our marketplace needs to
\begin{inparaenum}[(1)]
	\item enable the customer to make an informed decision about which solution mappings to include into the allocation and
	\item decide how much money has to be paid to each provider.
\end{inparaenum}

We use the following definition throughout the paper to denote single solution mappings, query answers, and allocations:
\begin{definition}[Solution Mapping and Query Answer]
	We denote as $\Omega$ the set of all possible solution mappings. We denote as $\omega\in\Omega$ a single solution mapping. A \emph{query answer} $\rho\subseteq\Omega$ is a set of solution mappings.
\end{definition}

\begin{definition}[Allocation]
	An \emph{allocation} $a\subseteq\rho$ is a set of solution mappings which are chosen from the query answer $\rho$. 
\end{definition}

\begin{figure}[ht]
	\centering
	\includegraphics[width=1.0\linewidth]{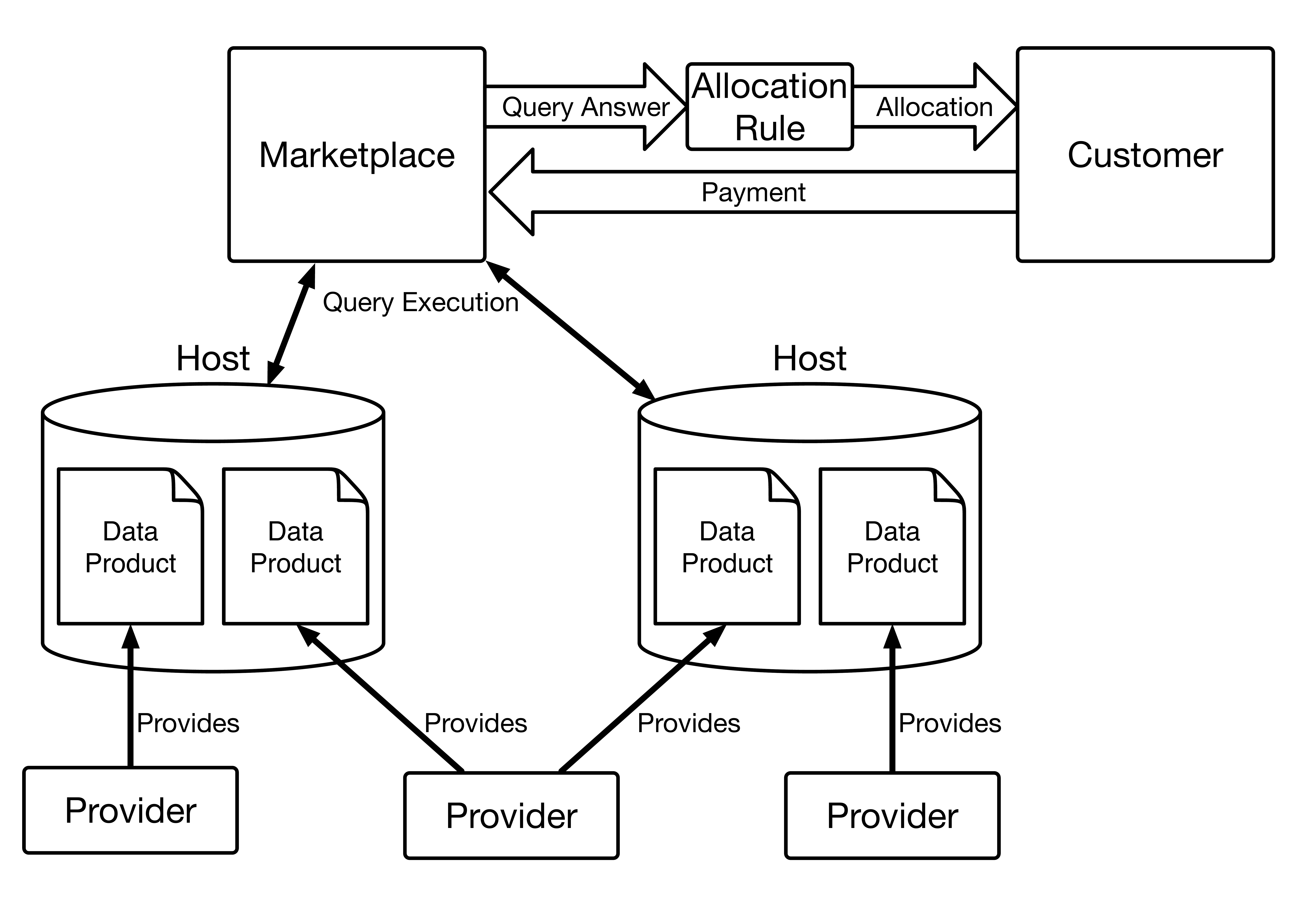}
	\caption{Customer, Host, and Provider in the marketplace.}\label{fig:customerHostProvider}
\end{figure}

We now introduce the four different entities our market concept brings together, \emph{providers}, \emph{hosts}, \emph{customers} and the \emph{marketplace}, all depicted in Fig. \ref{fig:customerHostProvider}.
In the following, we further elaborate on these entities before introducing an operationalization of our marketplace in Section \ref{sec:operationalization}.

\subsection{Provider}\label{sec:provider}

A provider is the originator of some data, which is used in the production of a query answer. Providers can group their data into different \emph{data products} having different prices \emph{per triple}. On the data level, a data product is just a collection of RDF-graphs, which are accessible under the same license. In addition to the price, data providers can specify other meta-data that might be relevant for customers such as terms\slash conditions of access, recentness, or origin of the underlying raw data.
Providers are responsible for the \emph{quality} of data, including \emph{recentness}, \emph{consistency} and \emph{accuracy}~\cite{serviceOrientedIntegration}.

\begin{definition}[Data Products and Prices]
	We will denote data products with $P_{1},\ldots,P_{k}$ and their price with $\pi_{1},\ldots,\pi_{k}$. Each Product $P_{i}$ consists of a collection $G_{i}$ of RDF graphs.
\end{definition}

The providers set the optimal price based on market conditions. This price can be learned, e.g., using reinforcement learning.
The price $\pi_{1}$ of the data product $P_{i}$ indicates the payment that is involved when using an \emph{RDF-Triple} contained in the data product for an allocation. This is an important aspect of our market concept: A provider is only payed for those triples which are \emph{allocated}. Triples which are accessed during query execution but not allocated, do not receive a payment from the customer.
The reasons for \textsf{FedMark} to only charge for allocated triples are the following:
\begin{enumerate}[(1)]
	\item If customers would have to pay for all the accessed triples, allocated or not, they would ``loose'' some of the paid triples by operations like joins, projections, and filters. Hence, it could be more beneficial to omit filters and projections and perform (left) outer joins to get as much data as possible for the \emph{same} price. This would incentivize customers to obtain data they do not need and optimize the query for the best data yield for a given price. We do not believe that it is desirable to create a market where such meta-optimizations are required from a customer to maximize the utility they can get out of our market.
	\item If providers would be compensated for accessed triples, the same query could yield very different revenue for a seller depending on query execution. Especially in join-operations, constraints from previous joins can vastly reduce the data that has to be accessed. This means that providers whose data is accessed earlier during query execution have the tendency to sell more data. In extreme cases, the price of an allocation is mainly dominated by the provider who is accessed first which, in turn, also receives the majority of the revenue. We believe that this potential for very \emph{unbalanced} vending of data is not desirable.
	\item Compensating accessed triples would create wrong incentives for the data providers: they would be able to increase revenue by increasing the amount of accessed triples, even if those additional triples do not contribute in any way to the final query answer. Providers would basically be encouraged to produce as much \enquote{dead weight} data -- data which is not really useful but has to be accessed during query execution because their lack of any use is only discovered after query execution -- as possible to maximize revenue.
	\item Charging for accessed triples adds an additional layer of complexity for query planning and optimization. In particular, the marketplace would have to estimate how many triples have to be accessed to calculate the price of a specific allocation prior to query execution. However, as we have shown in \cite{grubenmann2017-ISWC}, such estimations can be very unreliable. A high discrepancy between estimated price and actual price can be very devastating for such a marketplace.
\end{enumerate}

Note that providers do not serve their data; this is done by separate entities, the hosts. The separation between host and provider allows for more flexible business models for data provision, as some providers might have an initial budget to create some data (e.g., government subsidies) but do not have the funds to cover the \emph{operating costs} for running a SPARQL endpoint or may have other reasons to outsource the actual data provision. Providers can decide to act at the same time as a host for their own and\slash or other provider's data. Nevertheless, the market distinguishes between the two different roles, provider and host, and treats them as separate entities.

Data providers rely on the hosts to make their data available to the marketplace and thus, enable customers to buy their data. Similar to a Webhost for traditional Web content, the hosts in our market concept are paid by the provider based on some service agreement. Hence, the providers have to include the hosting costs into their pricing decision.

Similar to other digital goods such as software, eBooks, or digital music, the customer does not buy the good itself but buys the \emph{right} to use it under certain terms. For example, most usage rights for digital goods do not allow their resale to third parties. It is, however, the task of the provider to specify the exact terms under which the specific good is sold.

We elaborate on the role of providers in the following example:

\begin{example}\label{ex:providersExample}
	Data providers who want to contribute data to the query introduced in Example \ref{ex:ipa_query} must offer data products which include data about suggestions to the programming problems or success ratings of those suggestions. Every query answer that requires data from one or several such data products results in some payments for the data providers.
	
	Consider a data product $P_{A}$ with a price of $\$0.10$ per triple providing success ratings for suggestions. This means that $P_{A}$ can offer triples matching the first triple pattern in Listing \ref{lst:errorQuery}. $P_{A}$ is basically running a service where users are reporting on the failure or success of certain suggestions. Consider further two data products $P_{B}$ and $P_{C}$ with prices of $\$0.02$ and $\$0.03$ per triple, respectively, providing a database with the actual suggestions to various problems. $P_{B}$ and $P_{C}$ both can offer triples matching all but the first triple pattern in Listing \ref{lst:errorQuery}.
	
	In this example, we assume that $P_{B}$ and $P_{C}$ do not have overlapping data regarding suggestions. However, $P_{A}$ has overlapping data with both $P_{B}$ and $P_{C}$, which means $P_{A}$ provides success ratings to the suggestions provided by $P_{B}$ and $P_{C}$. This means that there are two different ways how a solution mapping can be obtained to the query in Listing \ref{lst:errorQuery}. Either the data from $P_{A}$ is joined with the data from $P_{B}$, in which case the solution mapping would cost \$0.18 (1 triple from $P_{A}$ for \$0.10 and 4 triples from $P_{B}$ for \$0.02), or the data from $P_{A}$ is joined with the data from $P_{C}$, in which case the solution mapping would cost \$0.22 (1 triple from $P_{A}$ for \$0.10 and 4 triples from $P_{B}$ for \$0.03).
	
\end{example}

\subsection{Host}\label{sec:host}

Hosts operate computers that run SPARQL endpoints for querying data products. They provide the computational and network resources needed to query the providers' data products. Hence, they ensure the reliability, availability, security, and performance, which are usually specified as \emph{Quality of Service}~\cite{serviceOrientedIntegration}. 

Like cloud service providers, hosts incur the fixed cost of operating the infrastructure, possibly some variable cost relative in the size of the data they store, and some marginal cost in form of the computational resources involved for each query they execute. The host's marginal costs occur whenever the providers' data are queried, independently of whether any data product will eventually get allocated or not. Similar to a Webhost for traditional Web content, the hosts in our market concept have to charge the data providers to cover their costs and make some profit.

Note that a host can store data from multiple data providers and that some data providers may choose to act as their own host. A host has to make sure that nobody can access the data without agreeing to the terms defined by the providers.

\subsection{Customer and Allocation Rule}\label{sec:allocation}\label{sec:customer}

A customer is a person, or a program acting on behalf of a person, who has a SPARQL \emph{query} and wants to buy an \emph{allocation} of solution mappings to this query. Depending on the marketplace, the customer might have the choice to use his or her own allocation rule, use one of the marketplace's built-in allocation rules, or use an allocation rule provided by a third party.

The allocation rule sits between the marketplace and the customer. Conceptually, the allocation rule is an independent entity which takes a query answer as input and produces an allocation as output. Practically, the allocation rule can be
\begin{inparaenum}[(1)]
	\item a part of the marketplace, in which case the user has to provide the marketplace with the necessary parameters for the specific allocation rule to run the allocation process,
	\item a part of the the customer, in which case the customer has to inform the marketplace about the chosen allocation,
	\item or, an independent entity, in which case this entity has to get the necessary parameters for the specific allocation rule from the customer, has to inform the marketplace about the allocation decision, and has to forward the allocation to the customer once the payments are done.
\end{inparaenum}
The third option is in particular useful if the allocation process is computationally expensive.

The marketplace will not deliver the full query answer to the allocation rule but will anonymize the query answer such that the allocation process has enough information to choose an allocation. Once the allocation rule informs the marketplace about the chosen allocation and the payments are done, the marketplace will deliver the actual data. In Section \ref{sec:implementation} we will discuss the process of anonymization and in Section \ref{sec:operationalization} will discuss the allocation process in detail.

\subsection{Marketplace}\label{sec:market}

The role of the marketplace is to coordinate the exchanges between the customers posing queries and the hosts serving answers based on the providers' data products. As such it can be seen as an extension of a traditional federated query engine with economic considerations.

The marketplace allows a customer to buy an allocation made from the data providers' triples. For this, the market needs to determine the solution mappings which can be allocated. Based on our previous work in \cite{grubenmann2017-ISWC}, it is unlikely that an allocation based on some data synopses will produce satisfying results for the customer. Hence, the marketplace has to execute the customer's query to create the solution mappings which could be potentially allocated. The marketplace can either run the query on all, for the query relevant, data products or rely on some source selection and join-prediction service (see \cite{saleem2014fine} for a survey) to preselect a set of the most promising data products.

The customer's payment for an allocation is independent of the query execution. Consequently, the marketplace can optimize the query execution based on traditional federated query optimization techniques without having to consider the prices of the different data products. After the query execution, the customer has to decide which of the obtained solution mappings to buy. Either the customer decides based on an allocation rule or directly chooses a set of solution mappings. This means that the market might execute the query on some data products' triples which might not be included in the customer's allocation, eventually. The customer has to pay the price for all allocated solution mappings to the marketplace, which redirects the money to the respective provider.

\begin{figure}[h]
	\centering
	\includegraphics[width=0.9\linewidth]{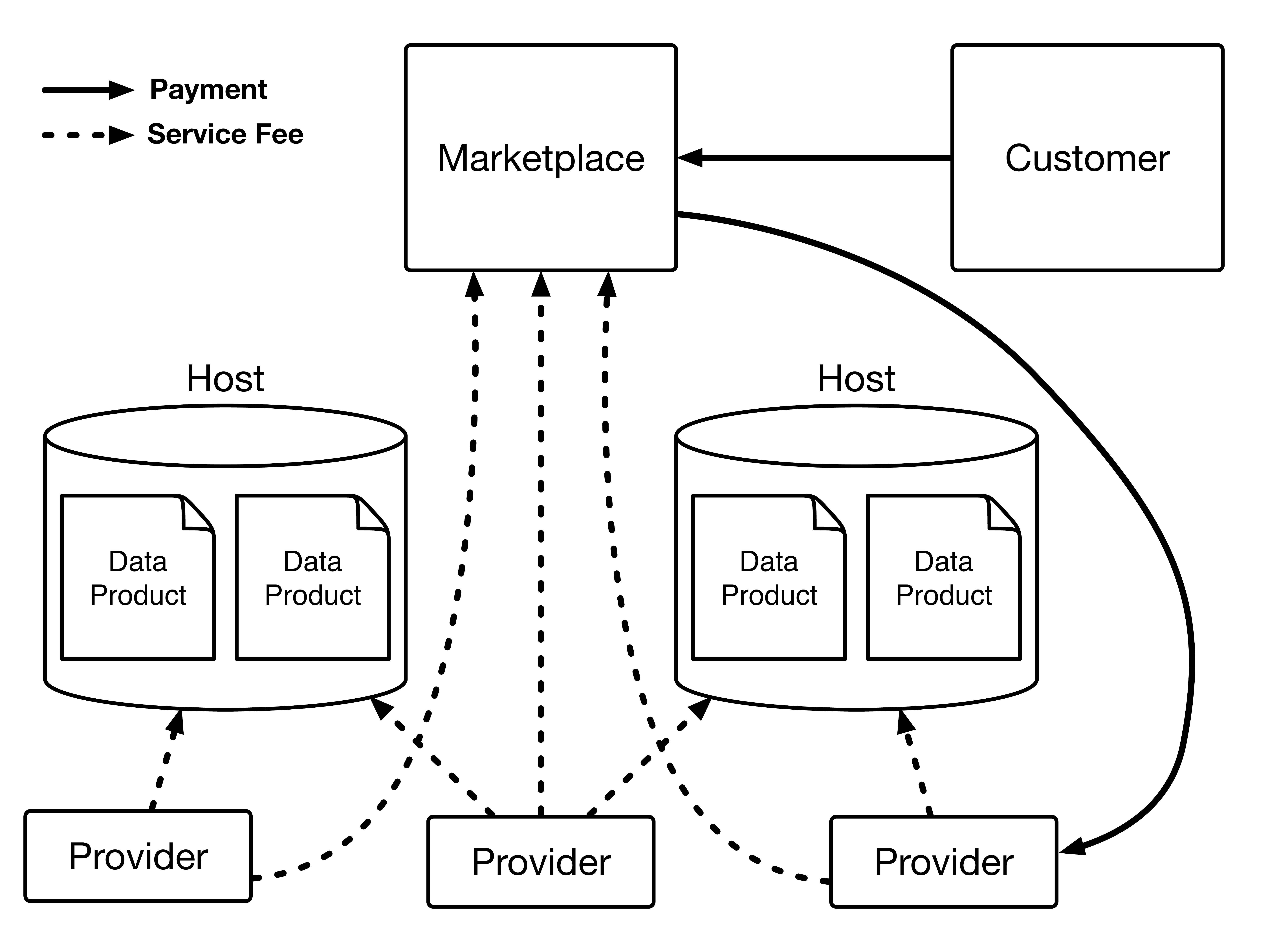}
	\caption{The customer pays the marketplace which forwards the money to the providers. The providers pay the marketplace and the hosts a certain fee for their services.}\label{fig:moneyflow}
\end{figure}

As discussed before, the providers will pay the hosts for their services. In addition, the providers also have to pay the marketplace a certain fee to keep it operational. Since the hosts and the marketplace are financially compensated by the providers, the providers will include these payments into their pricing decision. In addition, the market can use part of the generated revenue to subsidize providers which did not get allocated. However, if a provider fails to get allocated over a longer time, the provider's data might simply not be relevant at all, and the market can decide to stop subsidizing such providers. 
The payment to the host and the payment to the marketplace are transparent to the customer. Hence, the customer's allocation rule has to consider only the prices indicated by the data providers and not any additional payments to the hosts or market. Fig. \ref{fig:moneyflow} illustrates the money flow between marketplace, providers, hosts, and the customer. The solid arrows indicate payments which are inflicted whenever an allocation is delivered to a customer. The customer pays the marketplace which in turn pays the contributing providers. Note that in Fig. \ref{fig:moneyflow}, only one provider is contributing to the allocation and hence, only this provider is payed. All the providers in the marketplace have to pay service fees to the marketplace and the hosts, indicated by the dashed arrows, which are payed independent of the payment from the customer to the marketplace and to the providers.

\section{Implementation: FedMark}\label{sec:implementation}

In this section, we present our implementation of \textsf{FedMark}. \textsf{FedMark} is based on the federated querying engine FedX \cite{SchwarteHHSS11}. The core idea of \textsf{FedMark} is that each data product is represented as RDF statements, which describe the RDF data the product contains and any meta-information about the data. This allows one to access the necessary information about all data products as well as their contents with a single, federated SPARQL query. Additionally, it is possible to restrict the query answer including only providers having specific properties by changing the query, accordingly. We will now show how a SPARQL query can be rewritten to
\begin{inparaenum}[(1)]
	\item extract the necessary additional information about the product and
	\item exploit the information about a product to restrict the query answer.
\end{inparaenum}

In traditional federated SPARQL query execution, a SPARQL query is split up into subqueries which are sent to different endpoints. Combining the (sub-)query answers from the different subqueries results in the final query answer. Instead of sending the subqueries directly to the endpoints, \textsf{FedMark} replaces each occurrence of a triple pattern inside a subquery by a more complex graph pattern. This new graph pattern encloses the triple pattern into a \texttt{GRAPH} graph pattern. A \texttt{GRAPH} graph pattern uses the \texttt{GRAPH}-keyword to refer to the (named) RDF-graph which contains the matched triple pattern. \textsf{FedMark} also adds another triple pattern to refer to the product which contains the RDF-graph. Listing \ref{lst:graphGraphPattern} shows the general form of such a graph pattern. The \texttt{?graph} variable will be bound to the name (URI) of the graph which contains the triple matching the original triple pattern. If a product contains a certain RDF graph, this is expressed by the statement \texttt{?product market:contains ?graph}, where \texttt{?product} will be bound to the URI of the product. By referring to the URI of the product by using the variable \texttt{?product}, one can extract further information, e.g. the price, from the data product or restrict the query answer, e.g. by allowing only products having a rating greater than 8.0 (Listing \ref{lst:additionalProductInfos}). 

\begin{lstlisting}[basicstyle=\ttfamily,  
	morekeywords={GRAPH}, 
	caption={A Graph Pattern used for the execution of subqueries in \textsf{FedMark}.},label={lst:graphGraphPattern}]
GRAPH ?graph {
  [original triple pattern]
}
?product market:contains ?graph .
\end{lstlisting}

\begin{lstlisting}[basicstyle=\ttfamily,  
	morekeywords={GRAPH}, 
	caption={Additional information about a product are extracted and filtered.},label={lst:additionalProductInfos}]
?product market:price_usd ?price .
?product market:rating ?rating . FILTER (?rating >= 8)
\end{lstlisting}

Once the query is rewritten as described above, \textsf{FedMark} can execute the new subqueries on the available endpoints and create the query answer. This query answer is the basis on which the allocation rules can now decide which solution mappings to allocate. To prevent revealing the actual data before payment, \textsf{FedMark} does not give the actual query answer to the allocation rule. Instead, \textsf{FedMark} anonymizes the triples used to form the different solution mappings and reveals to the allocation a \emph{summary} which contains
\begin{inparaenum}[(1)]
	\item the anonymized triples which are needed to form a specific solution mapping and
	\item all meta-data available for these anonymized triples.
\end{inparaenum}

Table \ref{tab:anonymQueryAnswer} shows an example of such a summary. Each row in the left table represents one solution mapping of the query answer. The right table illustrates potential meta-data which could be available for the triples.
\begin{table}[htb]
\centering
\caption{Example Summary for the Allocation Rule}\label{tab:anonymQueryAnswer}
	\begin{tabular}{ll}
		\hline
		Solution Mapping & Triples  \\ \hline
		1 & $t_{1},t_{2}$ \\
		2 & $t_{1},t_{3}$ \\
		3 & $t_{2},t_{3}$ \\ \hline
	\end{tabular}
\quad
	\begin{tabular}{llc}
		\hline
		Triple & Price & Quality \\ \hline
		$t_{1}$ & $\$0.01$ & $\star$\\
		$t_{2}$ & $\$0.02$ & $\star$ $\star$ $\star$\\
		$t_{3}$ & $\$0.03$ & $\star$ $\star$\\ \hline
	\end{tabular}
\end{table}

\section{Allocation Rules}\label{sec:operationalization}

As introduced in the last section, a customer can use an \emph{allocation rule} to instruct a program which solution mappings should be allocated and returned. In contrast to a manual selection, an allocation rule allows a customer to automatize the whole process of buying an allocation. This possible automatization is an important aspect of a machine processable WoD, as it allows a customer to instruct a program to buy and process semantic data as needed without any interference from the customer.

The allocation rule is not part of the core concept of \textsf{FedMark}, as our marketplace does not have to know \emph{how} the customer decided for a specific allocation. The only important information is \emph{which} solution mappings are allocated by the customer. Hence, the allocation rule is transparent to \textsf{FedMark} and only its outcome is important. However, implementations of our \textsf{FedMark} concept can provide helpful interfaces and predefined allocation rules to support customers with formulating and implementing an appropriate allocation rule. Eventually, it is the responsibility of the customer to make a good allocation decision or come up with a good allocation rule to benefit most from the data \textsf{FedMark} can offer.

In the following, we want to present different allocation rules which could be implemented by the customer. As mentioned before, \textsf{FedMark} does not natively provide or constraint the allocation rule. Hence, the presented allocation rules are just a selection of possible allocation rules. We decided to discuss these allocation rules because they illustrate an interesting trade-off between optimality of the allocation and scalability with respect to the number of available solution mappings.

All allocation rules which we will present here have in common that they try to find an allocation which maximizes the customer's \emph{utility}. We assume that the utility is \emph{quasilinear}. This means that the utility of an allocation is the \emph{value} a customer has for this specific allocation minus the \emph{price} the customer has to pay for it. The customer's value indicates how much the customer is \emph{maximally} willing to pay for a specific allocation.

The price of an allocation is just the sum of the prices for each triple, as indicated by the data providers. The value of an allocation, however, is a private knowledge of the customer and needs to be defined with a function,  the \emph{valuation}. The valuation is used by the allocation rule to assert the value of a specific allocation. In the following, we will restrict ourselves to valuations which are \emph{linear} with respect to the solution mappings, this means that the customer's \emph{value} for an allocation is the sum of the values of the solution mappings contained in the allocation. The valuation is used to \emph{discriminate} between different possible allocations containing different solution mappings. 

\begin{definition}[Linear Valuation]
	A \emph{linear valuation} is a linear function $V:\mathcal{P}(\Omega)\rightarrow\mathbb{R}_{+}$ that assigns a value to each allocation $a\subseteq\rho$. The valuation has the form $a=\{\omega_{1},\ldots,\omega_{n}\}\mapsto\sum\limits_{i=1}^{n}v(\omega_{i})$, where $v(\omega_{i})=:v_{i}$ is the customer's value for solution mapping $\omega_{i}$.
\end{definition}

\begin{definition}[Customer's Utility]
	The customer's \emph{utility} $u\in\mathbb{R}$ for an allocation $a$ is the difference between the customer's value $v=V(a)$ of the allocation minus its \emph{payment} $\Pi(a)$, where $\Pi:\mathcal{P}(\Omega)\rightarrow\mathbb{R}_{+}$ is a function defined by the marketplace that determines the customer's payment for the allocation.
\end{definition}

In addition to the valuation, the customer has the possibility to add a \emph{budget constraint}. A budget constraint acts as a cap on the payment for the customer and allows the customer control over the maximal amount spent for an allocation.

Returning to our example, we include the customer's valuation:

\begin{example}\label{ex:value}
	A customer might be willing to pay up to $\$0.25$ for any solution mapping to the query in Listing \ref{lst:errorQuery}, but is willing to add an additional $\$0.10$ if all triples originate from a reliable source. In this case, every solution mapping to the query in Listing \ref{lst:errorQuery} has a value of $\$0.25$, if at least one of the sources is not considered reliable by the user, and a value of $\$0.35$, if all sources are considered reliable.
\end{example}

In Section \ref{sec:evaluation} we will compare different allocation rules and show under which circumstances which of them should be preferred.

\subsection{Integer Programming Allocation Rule}

The Integer Programming Allocation Rule maximizes the customer's \emph{utility} given a customer's query $q$, the valuation function $V(\cdot)$, the prices $\pi_{1},\ldots,\pi_{n}$, and the budget constraint $s$. Hence, the allocation rule describes an optimization problem. We will now show how we can express this optimization problem as an Integer Programming Problem: 

Let $\tau_{j}\in\{0, 1\}$ with $j\in\{1,\ldots,n\}$ be a binary variable indicating whether the triple $t_{j}$ is bought, $\pi_{j}$ the price associated with buying the triple, $r_{i}\in\{0, 1\}$ with $i\in\{1,\ldots,k\}$ a binary variable indicating whether the solution mapping $\omega_{i}$ can be obtained from the current allocation of triples, where $k$ is the number of all possible solution mappings, and $v_{i}:=v(\omega_{i})$ the value for the solution mapping $\omega_{i}$. Let further $s$ be the budget of the customer which acts as a cap on the total payment.

The objective is to find values for $\tau_{1},\ldots,\tau_{n}$ and $r_{1},\ldots,r_{k}$ which maximizes the utility $u(\tau_{1},\ldots,\tau_{n}, r_{1},\ldots,r_{k})$, that is the sum of the values of the allocated solution mappings minus the price of the necessary triples:
\begin{equation}
	u(\tau_{1},\ldots,\tau_{n}, r_{1},\ldots,r_{k}) = \left(\sum\limits_{j=1}^{k}r_{j}\cdot v_{j}\right) - \left(\sum\limits_{i=1}^{n}\tau_{i}\cdot \pi_{i}\right)
\end{equation}

In addition to the objective, we also have constraints which have to be respected by the solution of the Integer Programming Problem.

The first constraint is that for a solution mapping $\omega_{j}\in\Omega$ all the $n_{j}=|I_{j}|$ relevant triples $\{t_{i}~|~i\in I_{j}\}$  have to be allocated to include that solution mapping into the allocation, where $I_{j}$ is the \emph{index set} of the indices of the relevant triples. This means that the binary variable $r_{j}$ can only be set to one when all the variables $\tau_{i}$ with $i\in I_{j}$ are set to one. We can enforce this with the following linear constraint:
\begin{equation}
	\sum\limits_{i\in I_{j}}\tau_{i} - n_{j}\cdot r_{j}\geq 0 
\end{equation}

It is possible that multiple data products offer the same triple, in this case, only the cheapest triple will be considered. If multiple data products offer the same triple at the same price, one of the triples is randomly chosen as the relevant triple.
	
The second constraint is that the price of the allocation does not exceed the budget $s$:
\begin{equation}
	\sum\limits_{i=1}^{n}\tau_{i}\cdot \pi_{i}\leq s
\end{equation}

Equations \ref{eq:integerProgram} show the general form of the Integer Program for this optimization problem:
\begin{eqnarray}\label{eq:integerProgram}
	\text{Objective: }&\max \left(\sum\limits_{j=1}^{k}r_{j}\cdot v_{j} - \sum\limits_{i=1}^{n}\tau_{i}\cdot \pi_{i}\right)\\ \notag
	\text{Subject to: }
	&\sum\limits_{i\in I_{1}}\tau_{i} - n_{1}\cdot r_{1}\geq 0\\ \notag
	&\vdots \\ \notag
	&\sum\limits_{i\in I_{k}}\tau_{i} - n_{k}\cdot r_{k}\geq 0\\ \notag
	&\sum\limits_{i=1}^{n}\tau_{i}\cdot \pi_{i}\leq s\\ \notag
	\text{Bounds: }& r_{1},\ldots,r_{k}\in\{0,1\}\\ \notag
	& \tau_{1},\ldots,\tau_{n}\in\{0, 1\}\\ \notag
\end{eqnarray} 
	
In Example \ref{ex:integerProgram} we show how such an Integer Programming Problem for our scenario could look like.
	
\begin{example}\label{ex:integerProgram}
	We extend Example \ref{ex:value} by assuming that provider $P_{A}$ can contribute triples $t_{1},\ldots,t_{5}$, provider $P_{B}$ triples $t_{6},\ldots,t_{17}$, and provider $P_{C}$ triples $t_{18},\ldots,t_{25}$. Further, assume that the query answer consists of the solution mappings with their respective value according to Table \ref{tab:solutionsAndValues}.

	\begin{table}[htb]
	\centering
	\caption{Solution mappings, their required triples, and the value.}\label{tab:solutionsAndValues}
	\begin{tabular}{llll}
		\hline
	
		Solution Mapping & Triples & Value \\ \hline
			$\rho_{1}$ & $t_{1},t_{6},t_{7},t_{8},t_{9}$ & $\$0.25$\\
			$\rho_{2}$ & $t_{2},t_{10},t_{11},t_{12},t_{14}$ & $\$0.35$\\
			$\rho_{3}$ & $t_{3},t_{14},t_{15},t_{16},t_{17}$ & $\$0.35$\\
			$\rho_{4}$ & $t_{4},t_{18},t_{19},t_{20},t_{21}$ & $\$0.35$\\
			$\rho_{5}$ & $t_{5},t_{22},t_{23},t_{24},t_{25}$ & $\$0.25$\\ \hline
	\end{tabular}
	\end{table}
	
	In this case, the Integer Programming Problem has the following form:
	\begin{eqnarray}\label{eq:integerProgramExample}
		\text{Objective: }&\max (\$0.25 \cdot (r_{1}+r_{5})\\ \notag
	    &+~\$0.35 \cdot (r_{2} + r_{3} + r_{4}) \\ \notag
	    &-~\$0.10\cdot (\tau_{1} + \cdots + \tau_{5})\\ \notag
	    &-~\$0.02\cdot (\tau_{6} + \cdots + \tau_{17})\\ \notag
	    &-~\$0.03\cdot (\tau_{18} + \cdots + \tau_{25}))\\ \notag
		\text{Subject to: }
	    &\tau_{1} + \tau_{6} + \tau_{7} + \tau_{8} + \tau_{9} - 5\cdot r_{1} \geq 0\\ \notag
	    &\tau_{2} + \tau_{10} + \tau_{11} + \tau_{12} + \tau_{13} - 5\cdot r_{2} \geq 0\\ \notag
	    &\tau_{3} + \tau_{14} + \tau_{15} + \tau_{16} + \tau_{17} - 5\cdot r_{3} \geq 0\\ \notag
	    &\tau_{4} + \tau_{18} + \tau_{19} + \tau_{20} + \tau_{21} - 5\cdot r_{4} \geq 0\\ \notag
	    &\tau_{5} + \tau_{22} + \tau_{23} + \tau_{24} + \tau_{25} - 5\cdot r_{5} \geq 0\\ \notag
		&\$0.10\cdot (\tau_{1} + \cdots + \tau_{5})\\ \notag
		&+~\$0.02\cdot (\tau_{6} + \cdots + \tau_{17})\\ \notag
		&+~\$0.03\cdot (\tau_{18} + \cdots + \tau_{25})\leq \$0.65\\ \notag
		\text{Bounds: }& r_{1},\ldots,r_{5}\in\{0,1\}\\ \notag
		& \tau_{1},\ldots,\tau_{25}\in\{0, 1\}\\ \notag
	\end{eqnarray}
\end{example}

The two big advantages of the Integer Programming Allocation Rule are that
\begin{inparaenum}[(1)]
	\item it can be solved using standard optimization tools which are specialized in such problems and
	\item the allocation found by this rule is optimal: a customer cannot gain more utility by any other allocation, given the valuation and prices.
\end{inparaenum}
However, the allocation rule also has also a disadvantage:
Solving the Integer Programming Problem is NP-hard, which means that the Integer Programming Allocation Rule has a limited scalability. We will investigate the runtime needed to solve the problem in Section \ref{sec:evaluation}. Another drawback of this allocation rule is that it is limited to a linear valuation of the allocation. As soon as the value for a single solution mapping is not constant---this can happen for example if the customer has a decreasing marginal value for the solution mappings---the optimization cannot anymore be formulated as an Integer Programming Problem and the solving tools cannot be used.
Given this drawback, we present another allocation rule which will complement the Integer Programming Allocation Rule.
		
\subsection{Greedy Allocation Rule}

The Greedy Allocation Rule tries, as the name suggests, to find a good allocation in a greedy fashion. Hence, the allocation rule does not guarantee that the found allocation is optimal in the sense that it maximizes the customer's utility. The upside of this allocation rule is that
\begin{inparaenum}[(1)]
	\item it scales better with increasing number of solution mappings which can be allocated, which means it remains feasible in situations where the NP-hard Integer Programming Allocation Rule is not anymore feasible, and
	\item the allocation rule is compatible with any valuation which is monotonic decreasing.
\end{inparaenum}
Note that the valuation does not have to be strictly monotonic decreasing but can also be linear.

The idea of the Greedy Allocation Rule is quite simple: Choose the non-allocated solution mapping with the highest \emph{ratio between utility and price} and allocate it, as long as the utility is positive (the value is higher than the price) and the sum of the price of all allocated solution mappings is smaller than or equal to the budget. Whenever a new solution mapping is allocated, the utility of the remaining non-allocated solution mappings must be updated. This is because an allocated solution mapping might include some triples of the non-allocated solution mappings, in which case the specific triples do not have to be bought again and the price for the respective solution mappings decreases.

Fig. \ref{alg:greedy} shows how the Greedy Allocation rule can be implemented. First, the algorithm initializes the set $I$ with the indices of all solution mappings, the set $T_{\textnormal{buy}}$ for the indices of all triples which need to be bought, the allocation $a$, and the total price $\Pi$ of allocation $a$ (Line \ref{alg:greedy:initialization_start}--\ref{alg:greedy:initialization_stop}). Then, the algorithm enters a loop and determines the solution mapping with the highest ratio between utility and price. For this, the algorithm has to get the indices of those triples which are relevant for a specific solution mapping $\omega_{i}$. This information is provided by the function \emph{relevantTriplesIndices} (Line \ref{alg:greedy:getRelevantTriples}). Afterwards, the indices of those triples which are already considered for buying are removed from $T_{\textnormal{buy}}$ (Line \ref{alg:greedy:removeAllocatedTriples}), because the same triple does not need to be bought twice by the customer. Using the indices of the required triples, the price $\Pi_{i}$ (Line \ref{alg:greedy:price}), the utility $u_{i}$ (Line \ref{alg:greedy:utility}), and the ratio $r_{i}$ (Line \ref{alg:greedy:ratio}) can be calculated. With this information, the algorithm can determine the index $max$ of the solution mapping with the highest ratio having a price which is still in the budget $b$ and a utility of at least 0 (Line \ref{alg:greedy:max}). If there is still a solution mapping that reaches these conditions (Line \ref{alg:greedy:if}), the total price $\Pi$, the allocation $a$, the index set $I$ with all available solution mappings for allocation, and the index set of all triples to be bought $T_{\textnormal{buy}}$ are updated accordingly (Line \ref{alg:greedy:update_start}--\ref{alg:greedy:update_stop}). If there is no suitable solution mapping or simply no solution mapping at all left, the algorithm stops the loop and returns the allocation $a$ and the total price $\Pi$ (Line \ref{alg:greedy:condition}--\ref{alg:greedy:return}).

\begin{figure}
	\begin{algorithmic}[1]
		\Require{Solution mappings $\omega_{1},\ldots,\omega_{n}$, values $v_{1},\ldots,v_{n}$, prices $\pi_{1},\ldots,\pi_{k}$, budget $b$}
		\Ensure{Allocation $a$, Payment $\Pi$}
		\State $I$ $\leftarrow$ $\{1,\ldots,n\}$\;\label{alg:greedy:initialization_start}
 		\State $T_{\textnormal{buy}}$ $\leftarrow$ $\emptyset$\;
 		\State $a$ $\leftarrow$ $\emptyset$\;
 		\State $\Pi$ $\leftarrow$ $0$\;\label{alg:greedy:initialization_stop}
 		
 		\Do
		  \For{$i\in I$}
		   \State $T_{\textnormal{relevant}}$ $\leftarrow$ $\textnormal{relevantTriplesIndices}(\omega_{i})$\;\label{alg:greedy:getRelevantTriples}
		   \State $T_{\textnormal{required}}$ $\leftarrow$ $T_{\textnormal{relevant}}\setminus T_{\textnormal{buy}}$\;\label{alg:greedy:removeAllocatedTriples}
		   \State $\Pi_{i}$ $\leftarrow$ $\sum\limits_{j\in T_{\textnormal{required}}}\pi_{j}$\;\label{alg:greedy:price}
		   \State $u_{i}$ $\leftarrow$ $v_{i}-\Pi_{i}$\;\label{alg:greedy:utility}
		   \State $r_{i}$ $\leftarrow$ $\frac{u_{i}}{\Pi_{i}}$\;\label{alg:greedy:ratio}
		  \EndFor
		  \State $max$ $\leftarrow$ $\textnormal{argmax}_{i\in\{1,\ldots,n\}}(\{r_{i}|\Pi+\Pi_{i}\leq b\textnormal{ and }u_{i}\geq0\})$\;\label{alg:greedy:max}
		  \If{$\exists max$}\label{alg:greedy:if}
		   \State $\Pi$ $\leftarrow$ $\Pi+\Pi_{max}$\;\label{alg:greedy:update_start}
		   \State $a$ $\leftarrow$ $a\cup\{\omega_{max}\}$\;
		   \State $I$ $\leftarrow$ $I\setminus\{i\}$\;
		   \State $T_{\textnormal{buy}}$ $\leftarrow$ $T_{\textnormal{buy}}\cup T_{\textnormal{relevant}}$\;\label{alg:greedy:update_stop}
		  \EndIf
		 \DoWhile{$\exists max$ and $I\neq\emptyset$}\label{alg:greedy:condition}
		 \State \Return $a$, $\Pi$ \label{alg:greedy:return}
	\end{algorithmic}
	\caption{Algorithm for the Greedy Allocation Rule.}\label{alg:greedy}
\end{figure}

The allocation rule runs in $O(n^{2}\log(n))$ time, where $n$ is the number of solution mappings available: For each (at most $n$) new allocated solution mapping, the algorithm has to sort the remaining non-allocated solution mappings ($O(n\log(n))$) to determine the one with the highest utility.

\section{Evaluation}\label{sec:evaluation}

To empirically compare the presented allocation rules, we measure runtime and utility for different queries. The first measure, runtime, is chosen to establish that the use of our proposed method is \emph{feasible} and practical in a WoD setting from a computational point of view. The second metric, utility, indicates that the results are \emph{desirable} and how close the greedy rule approaches the optimal.

\subsection{Evaluation Setup}
We use two different scenarios: one based on the FedBench benchmark and a new, synthetic scenario.

\textbf{FedBench Scenario:} The goal of this first scenario is to evaluate our procedure in a well-established realistic setting. FedBench \cite{Fedbench2011} consists of 9 datasets on various topics and 25 SPARQL queries. We excluded queries with only 1, 2, or 3 solution mappings, as their allocation would be rather trivial. This left us with 17 queries. The number of solution mappings range from 11 to 9054 per query. Table \ref{tab:queries} shows the number of solution mappings and triples per solution mapping for each of the 17 selected FedBench queries.

\begin{table}[tbp]
\centering
\caption{Number of solution mappings and triple per solution mapping.}
\label{tab:queries}
\begin{tabular}{lrc}
\hline
Query & Solution Mappings & Triples per Solution Mapping \\
\hline
CD1   & 90   & 1--2  \\
CD6   & 11   & 4     \\
LD1   & 308  & 3     \\
LD2   & 185  & 3     \\
LD3   & 159  & 4     \\
LD4   & 50   & 5     \\
LD5   & 28   & 3     \\
LD6   & 39   & 5     \\
LD7   & 1216 & 2     \\
LD8   & 22   & 5     \\
LD11  & 376  & 5     \\
LS1   & 1159 & 1     \\
LS2   & 333  & 1--2  \\
LS3   & 9054 & 5     \\
LS5   & 393	 & 6     \\
LS6   & 28   & 5     \\
LS7   & 144  & 4--5  \\
\hline
\end{tabular}
\end{table}


\textbf{Synthetic Scenario:} The goal of this second scenario is to evaluate the scaling behavior of the allocation procedure whilst varying both the number of solution mappings per query answer and the number of unique triples contained therein. To that end we generated hypothetical queries that have randomly generated query answers (as we only require the answer sets for evaluating the allocation procedure). The number of solution mappings $s$ per answer varies between 50 and 1000. Each solution mapping consists of 5 triples, hence we have $n=5s$ triples in an answer. To simulate the \emph{diversity} of an answer, we introduce the parameter $d \in [0,1]$, which specifies how many unique triples are contained in a query answer. Next, we randomly assign each triple one of $n_{\textnormal{unique}}=1 + d\cdot(n-1)$ identifiers. The result is an answer set in which all triples are the same for $d=0$ and each triple in the query answer is unique for $d=1$.
This procedure generates query answers of varying size $s$ and number of unique triples $n_{\textnormal{unique}}$. 

\textbf{Parameters:} We set the price for each triple to a uniformly distributed random number between $\$0.00$ and $\$1.00$. The number of triples per solution mapping multiplied by a uniformly distributed random number between $\$1.00$ and $\$2.00$ gives us the value for each solution mapping. The budget for each query is set such that only $50\%$ of the solution mappings can be obtained. Using these numbers ensures that
\begin{inparaenum}[(1)]
	\item there is at least one affordable allocation having positive utility and
	\item not all solution mappings can be allocated.
\end{inparaenum}
This guarantees that the allocation problem does not become trivial to solve. \\

\subsection{Results}
We discuss both scenarios in turn.
 
\begin{figure*}
	\centering
	\includegraphics[width=1.0\linewidth]{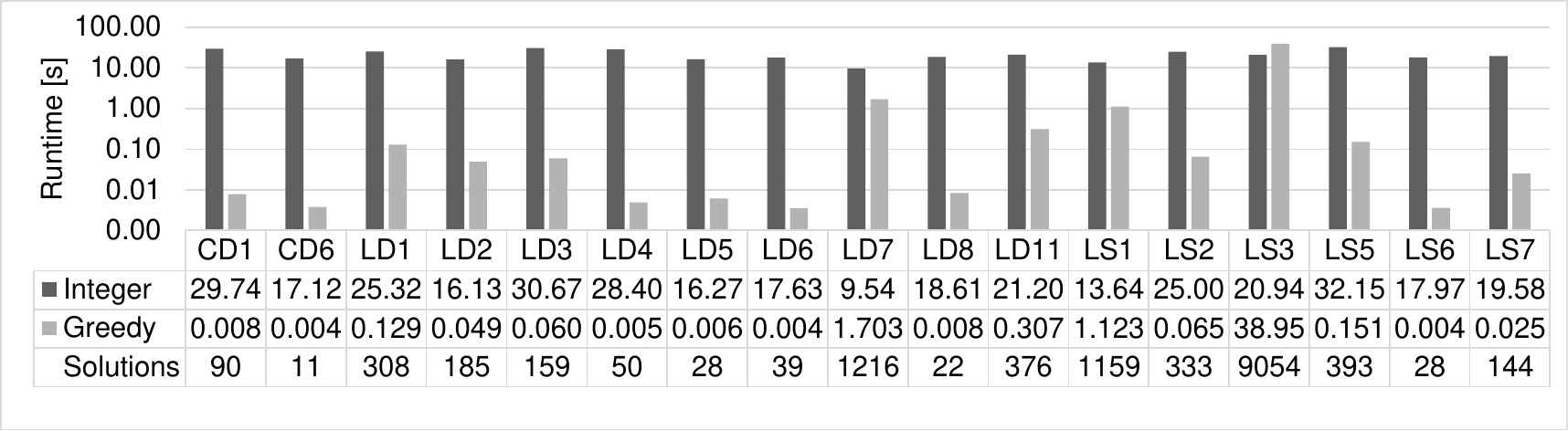}
	\caption{Runtime in seconds for the Integer Programming Allocation Rule (Integer) and the Greedy Allocation Rule (Greedy) for the FedBench benchmark.}\label{fig:runtime}
\end{figure*}

\textbf{FedBench Scenario:} Fig. \ref{fig:runtime} graphs the execution time for the 17 selected FedBench queries for the Integer Rule and the Greedy Rule in seconds. It shows that the Integer Allocation Rule is by orders magnitude slower than  the Greedy Allocation Rule for most queries. One exception is query LS3, which actually has a longer runtime for the Greedy Rule. LS3 is also the query with the highest number of solution mappings. One explanation is the high diversity in a large number of solution mappings that benefits the integer approach.

\begin{figure*}
	\centering
	\includegraphics[width=1.0\linewidth]{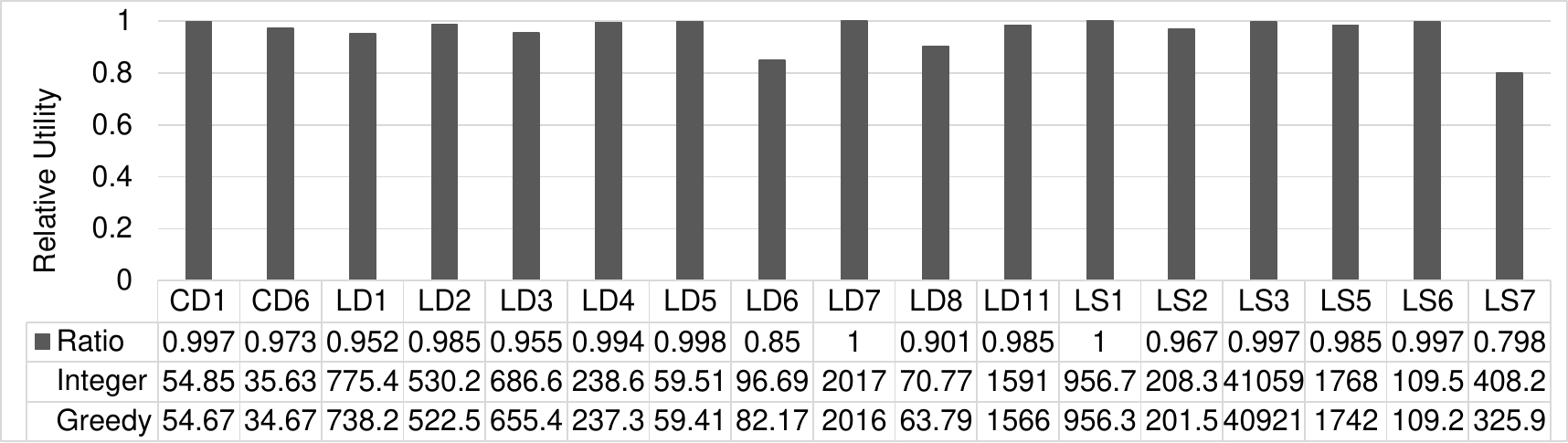}
	\caption{Ratio between the utility of the Greedy Allocation Rule (Greedy) and the Integer Programming Allocation Rule (Integer) for the FedBench benchmark.}\label{fig:utility}
\end{figure*}

The ratio between the utility of the Greedy Rule and the Integer Rule for the 17 selected FedBench queries is graphed in Fig. \ref{fig:utility}. The graph shows that the Greedy Rule has a utility which is very close to the Integer Rule, which maximizes utility given the prices and values. The evaluation shows that for the FedBench queries, the Greedy Rule provides allocations of comparable quality to the Integer Rule in orders of magnitude smaller time. 

\begin{figure*}
	\centering
	\includegraphics[width=1.0\linewidth]{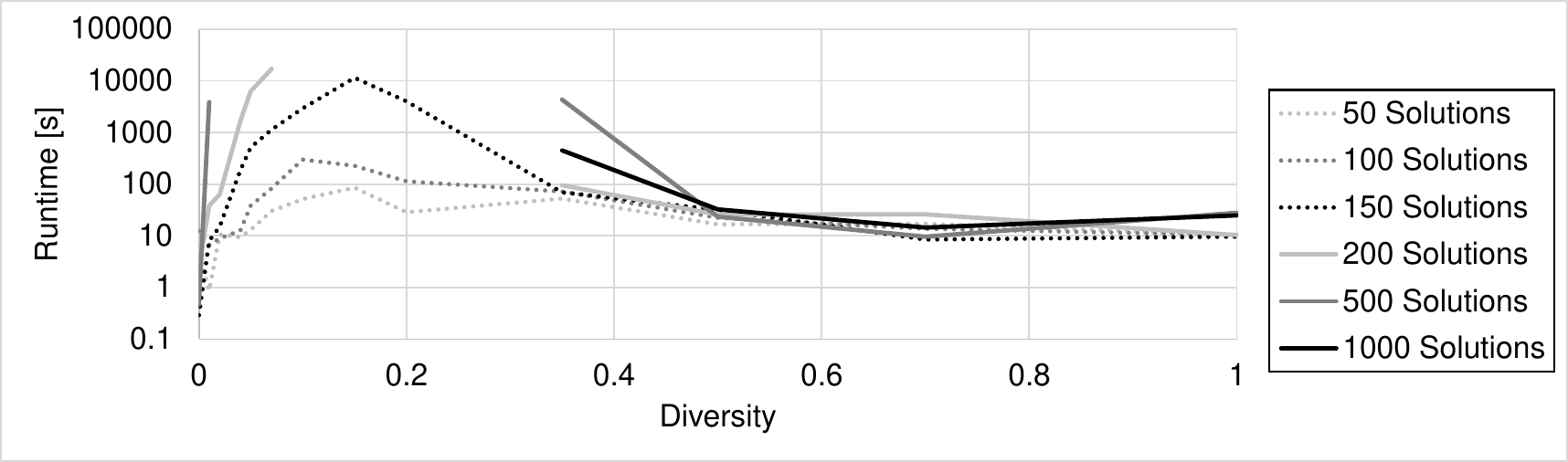}
	\caption{Runtime in seconds (on a logarithmic scale) for the Integer Programming Allocation Rule for different diversities and query answer sizes.}\label{fig:integerScalingRuntime}
\end{figure*}

\textbf{Synthetic Scenario:} We will now focus on the scaling behavior of both allocation rules. Fig. \ref{fig:integerScalingRuntime} shows how the Integer Programming Allocation Rule scales with respect to the diversity $d$ for different number of solution mappings. Note that the runtime is plotted in a logarithmic scale and that we plot with respect to $d$ and not $n_{\textnormal{unique}}$, as the latter is dependent on $s$. For this evaluation, we used our own synthetic data as described above. For some plots, the graph has some missing points. The missing points indicate parameter combinations that did not yield results within 12 hours of optimization. 
As the figure shows, the runtime complexity explodes if the diversity is in the lower third of the spectrum and the number of solution mappings is high enough. For a diversity of 0, the allocation problem becomes trivial as there is only one triple which can be chosen. For a diversity of 1, the different solution mappings in a query answer are \emph{independent}, meaning that they do not share any triples. Also, in this case, the allocation problem seems to be simpler to solve, although not as simple as when the diversity is 0. For a diversity in between 0 and 1, the allocation problem becomes harder. This is because the solution mappings are now more dependent on each other because they share triples: some combinations of solution mappings will be much cheaper than other combinations, because they can exploit the fact that they share some triples and their prices. Interestingly, the less the diversity is, the more time the Integer Programming Allocation Rule needs to solve the allocation problem. At least, until the diversity gets close to 0, at which point the number of triples is very low and the allocation problem becomes much easier.

\begin{figure*}
	\centering
	\includegraphics[width=1.0\linewidth]{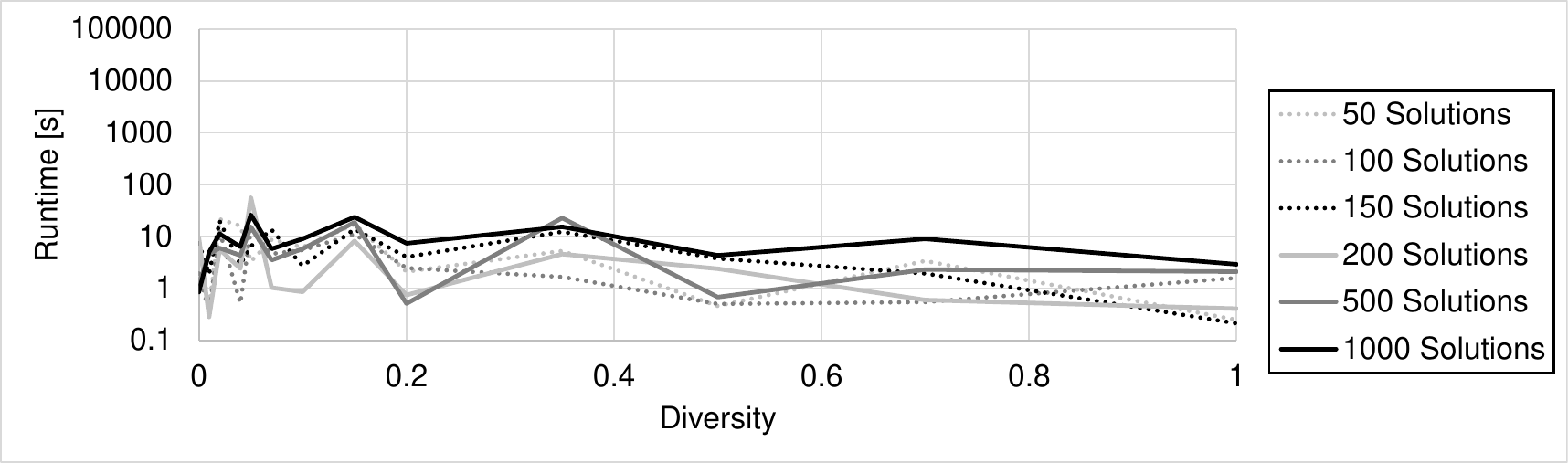}
	\caption{Runtime in seconds (on a logarithmic scale) for the Greedy Allocation Rule for different diversities and query answer sizes.}\label{fig:greedyScalingRuntime}
\end{figure*}

The scaling behavior of the Greedy Allocation Rule with respect to the diversity $d$ for different number of solution mappings is shown in Fig. \ref{fig:greedyScalingRuntime}. The plot uses the same scale as in Fig. \ref{fig:integerScalingRuntime} to make it easier to compare the results. As Fig. \ref{fig:greedyScalingRuntime} indicates, the runtime for the Greedy Allocation Rule does not suffer from the same explosion of runtime as the Integer Programming Allocation Rule when the diversity is low. The reason for this is quite simple: the Greedy Rule does have to consider which combination of solution mapping could exploit the overlap of triples the most. Instead, the allocation rule just chooses the next best solution mapping and updates the prices, accordingly. Nevertheless, we can observe some trend that the runtime is higher for lower diversity than it is for high diversities or a diversity of 0. This can be explained by considering how much the Greedy Allocation Rule has to resort the solution mappings after each step. If the diversity is high, there are only few solution mappings for which the price changes after a solution mapping is selected. Hence, the resorting can be done faster. If the diversity is low, however, selecting a solution mapping does impact more remaining solution mappings, due to the increased overlap. Hence, the resorting takes more time. Eventually, for a diversity of 0, after selecting the first solution mapping all other solution mappingss have a price of 0 (because they all need the same triple which is already bought), which means that no resorting at all is needed.\\

As our evaluation shows, the runtime of the Integer Rule depends highly on the diversity of the triples within a query answer. Hence, even a query answer with a lot of solution mappings can be feasible for the Integer Rule whereas another query answer with fewer solution mappings might not be feasible. The Greedy Rule behaves more stable with changing diversity. Paired with our observations about the utility of resulting solution mappings in the FedBench scenario, we can infer that the Greedy approach seems to provide good allocations within reasonable time bounds for realistic scenarios.

\section{Limitations and Conclusions}

To grow further and be able to serve as a high-quality data source, the WoD has to find the means to fund the creation, serving, and maintenance of data sources. In this paper, we proposed a new paradigm for funding these activities in the form of a market for data that combines a market-based approach with principles of federated querying. We presented \textsf{FedMark}, a prototype that implements the concepts we introduced in this paper. In addition, we introduced two possible allocation rules which can be used by a customer to decide for a specific allocation. As we have seen, both allocation rules have different properties with respect to runtime and utility. While the Integer Allocation Rule guarantees an optimal allocation with respect to utility, the runtime of this rule can exceed any reasonable time limit under certain condition, as we have seen in the evaluation section. The Greedy Allocation Rule does not suffer from extensive runtimes in the scenarios we investigated. However, the Greedy Rule cannot guarantee an optimal allocation. Although, we have seen that the utility is often very close to an optimal allocation. In practice, a customer would be advised to run both allocation rules in parallel and specify a time out. After the time runs out, the customer can check whether the Integer Allocation Rule has found an optimal allocation. If not, the current best solution of the Integer Allocation Rule can be compared to the outcome of the Greedy Rule. Whichever rule produces the best result under the time constraints should be picked by the customer.

Another advantage of the Greedy Rule is that it can handle decreasing marginal values for the solutions. If the customer's valuation is not linear, he might not have any other choice but to use the Greedy Rule. Obviously, we need to explore a set of other allocation rules to understand the trade-offs for the various different valuation needs of the customer.

In addition, we need to revisit our assumptions and check if they are truly realistic. 
As an example, consider the assumption that providers can amortize their fixed costs over many transactions. This is only true if their goods are actually sufficiently attractive to be bought, which again depends on the competitiveness of the marketplace. Whilst we believe that this is true for many data products (e.g., financial data) we will have to investigate where this assumption does not hold.
Second, this paper did not discuss how a provider decides on the optimal pricing. Whilst we did run an analysis indicating that it is favorable for a provider to learn the price, we did not evaluate how well that price \emph{can} be learned---a task for future work. 
Third, we need to explore the possibility of selling query subscriptions, which opens the way to mechanisms akin to the ones that are currently dominating the entertainment industry. 
Fourth, we need to explore market-aware optimizations for \textsf{FedMark} and evaluate their influence on the speed of query execution. 
Finally, the generalizability of our evaluation might be hampered by the use of FedBench. Indeed, FedBench's limitations led us to run a second evaluation with synthetic data. Whilst an evaluation in additional real-world scenarios is desirable and should be subject of future work, we believe that our evaluation highlights the key properties of our allocation rules and, hence, establishes their applicability.

Whatever the shortcomings of \textsf{FedMark} and our concept, we believe that the contributions of this paper are a first step in the principled exploration of a financially stable and, therefore, sustainable Web of Data.

\section*{Acknowledgments} This work was partially supported by the Swiss National Science Foundation under grant \#153598.

\end{document}